\documentclass[aps,prd,superscriptaddress,nofootinbib,amsmath,amsfonts,preprintnumbers,groupedaddress,showpacs,10pt,english]{revtex4}
\usepackage{amsmath}
\usepackage{amssymb}
\usepackage{babel}
\usepackage{wrapfig}
\usepackage{cancel}

\usepackage{relsize,exscale}
\makeatletter

\usepackage{array,multirow,graphicx}
\usepackage{dcolumn}
\usepackage{newlfont}
\usepackage{bm}
\usepackage[colorlinks,citecolor=blue,urlcolor=blue,linkcolor=blue]{hyperref}
\usepackage[figtopcap]{subfigure}
\usepackage{color}

\begin{document}
\tolerance=5000

\title{Charged spherically symmetric   black holes in scalar-tensor Gauss-Bonnet gravity}

\author{ Salvatore Capozziello }\email{ capozziello@na.infn.it, corresponding author.}
\affiliation{Dipartimento di Fisica ``E. Pancini``,  Universit\`a  di Napoli ``Federico II'',
Complesso Universitario di Monte Sant' Angelo, Edificio G, Via Cinthia, I-80126, Napoli, Italy, }
\affiliation{Istituto Nazionale di Fisica Nucleare (INFN),  Sezione di Napoli,
Complesso Universitario di Monte Sant'Angelo, Edificio G, Via Cinthia, I-80126, Napoli, Italy,}
\affiliation{Scuola Superiore Meridionale, Largo S. Marcellino 10, I-80138, Napoli, Italy.}

\author{ Gamal  G. L. Nashed}
\email{nashed@bue.edu.eg}
\affiliation{Centre for Theoretical Physics, The British University in Egypt, P.O. Box 43, El Sherouk City, Cairo 11837, Egypt,}

\date{\today}

\begin{abstract}
We derive a novel class of four-dimensional black hole solutions in Gauss-Bonnet gravity coupled with a scalar  field in presence of Maxwell electrodynamics. In order to derive such solutions, we assume  the ansatz   $ g_{tt}\neq g_{rr}{}^{-1}$ for metric potentials. Due to the  ansatz  for the metric, the Reissner Nordstr\"om gauge potential cannot be recovered because of the presence of higher-order terms  Â ${\cal O}\left(\frac{1}{r}\right)$ Â  which are not allowed to be vanishing. Â Moreover, the scalar field  is not allowed to vanish. If it vanishes, a function of the solution results undefined. Â 
Furthermore, it is possible to  show that the electric field is of higher-order in the monopole expansion:  this fact  explicitly comes from the contribution of the scalar field. Therefore, we can conclude that the Gauss-Bonnet scalar field acts as  non-linear electrodynamics  creating monopoles, quadrupoles, etc. in the metric potentials. We compute the invariants associated with the black holes and show that, when compared to Schwarzschild or Reissner-Nordstr\"om space-times, they have a soft singularity. Also, it is possible to demonstrate  that these black holes give rise to  three horizons in AdS space-time and  two horizons in dS space-time. Finally,  thermodynamic quantities  can be derived and we show that  the solution can be stable or unstable depending on  a  critical value of the temperature.
\end{abstract}


\maketitle
\section{Introduction}\label{S1}
A large amount of observational data indicates that our universe is experiencing an accelerated expansion. There are two basic approaches to explain this cosmic acceleration: Â The first considers the issue of acceleration in the framework of Einstein's general relativity (GR) and then needs the existence of an odd type of energy called "dark energy," which has a repulsive gravity and constitutes an unclustered ingredient  of  universe components. The second approach  proposes extensions   of GR by including functions of the   curvature invariants like the Ricci scalar  $R$, the Riemann and Ricci tensors or their derivatives in the Lagrangian formulation or modifications of the Einstein paradigm involving torsion tensor or non-metricity. Among these proposals to modify Einstein's theory there are  $f(R)$ gravity,  Â braneworld cosmology, Lovelock gravity,  Brans-Dicke-like   scalar-tensor theories,  etc. \cite{Brans:1961sx,Wagoner:1970vr,Barrabes:1997kk,Cai:1996pj,Capozziello:2005bu,Moffat:2005si,Dehghani:2006xt,Faraoni:2007yn,Maeda:2009js,Hendi:2012as,
Sharif:2013lea,Eiroa:2014mca,Karami:2014tsa,Darabi:2013caa,Akbar:2006mq,deSouza:2007zpn,Corda:2009bv,Cognola:2007zu,Nojiri:2010wj,Said:2011dg,Myung:2011ih,Hwang:2011kg,
Hendi:2011eg,Mazharimousavi:2011bf,Man:2013sf,Miao:2014wpa,Hendi:2012zg,Hendi:2014wsa,Hendi:2013zba,Ida:1999ui,Cline:2000xn,Brax:2003fv,Mizuno:2004xj,Nihei:2004xv,
Gergely:2006ei,Demetrian:2005sr,Lovelock:1971yv,Lovelock:1972vz,Deruelle:1989fj,Hendi:2015oda,Hendi:2015pda}.  Among these extensions a particular role can be that recently assumed by the so called Gauss-Bonnet (GB)
gravity where the topological invariant
\begin{equation}
\label{GBinv}
\mathcal{G}\equiv  R^2-4R^{\alpha\beta}R_{\alpha\beta}+R^{\alpha\beta\mu\nu}R_{\alpha\beta\mu\nu}\,,
\end{equation}
is considered into dynamics \cite{Bajardi,Paolella,Dialektopoulos,Sergey,Oikonomou:2015qha,Hendi:2016tiy,Hod:2019pmb,Ma:2019ybz,Ghosh:2019pwy,Odintsov:2020zkl,Odintsov:2020sqy,Blazquez-Salcedo:2020caw,Pozdeeva:2020shl,Kleihaus:2020qwo,Samart:2020mnn,Haroon:2020vpr,Parai:2020jrg,EslamPanah:2020hoj}. Specifically, the Lovelock theory is a well-known higher derivative  theory of gravity  that is a natural extension of  GR  where GB contributions are taken into account Â \cite{Stelle:1977ry,Maluf:1985fj,Farhoudi:2005qd}. Â Interestingly, there are no ghost terms if GB  curvature-squared terms are present, and the associated field equations  include only   metric's second derivatives \cite{Boulware:1985wk,Zumino:1985dp,Cho:2001su,Cai:2001dz,Myers:1987qx,Callan:1988hs}. Another interesting feature of GB gravity is the fact that it can arise from the low-energy limit ofÂ 
string heterotic theory \cite{Zwiebach:1985uq,Gross:1986iv,Gross:1986mw,Metsaev:1987bc,Metsaev:1987zx,Metsaev:1986yb}. It is found that the low-energy expansion of a closed heterotic string
effective action possesses a GB term and a scalar field \cite{Metsaev:1986yb}, however the functional form of the  scalar field is still a debated topic in view of some possible observational signature.

In the context of  GB gravity,Â   several black hole (BH) solutions  have been discussed Â in the literature \cite{Kim:1999dq,Cho:2002hq,Charmousis:2002rc,Kofinas:2003rz,Cai:2003gr,Barrau:2003tk,Maeda:2003vq,deRham:2006pe,Dotti:2007az,Brown:2006mh,Maeda:2007cb,Charmousis:2008kc,Hendi:2010zza,
Bouhmadi-Lopez:2013gqa,Hendi:2014bba,Yamashita:2014cra,Maselli:2015tta}.
In Ref. \cite{Li:2013cja,Zeng:2013fsa,Zhang:2014cga} holographic thermalization in GB gravity via the Wilson loops and holographic entanglement entropy has been studied.

If we are dealing with GR corrected by a GB term, we are considering the so-called \textit{Einstein-Gauss-Bonnet} (EGB) gravity. It has been demonstrated that  conserved charges of EGB- AdS gravity emerges in the electric section of the Weyl tensor: this feature  broadens the notion of conformal mass \cite{Jatkar:2015ffa}.
A four-dimensional Â EGB gravity possesses many interesting properties. It could, for example, solve some singularity issues. Â In particular, by taking into account four-dimensional EGB gravity and deriving a spherically symmetric BHs, the   field of gravity becomes repulsive as $r \to 0$, and therefore  infalling object cannot approach singularity. Â Additionally,  spherically symmetric  solutions in EGB  is distinguish  from  solutions of GR  like the Schwarzschild one. Additionally, investigations into compact objects and their physical characteristics have been conducted within the context of four-dimensional Einstein-Gauss-Bonnet (EGB) gravity. These studies encompass aspects such as black hole stability, quasi-normal modes, shadow phenomena, and the concept of strong cosmic censorship \cite{Konoplya:2020bxa,Mishra:2020gce,Roy:2020dyy}. A wide range of investigations has been carried out in the realm of gravitational physics. These studies encompass various aspects, including: The analysis of the shadow of a rotating black hole \cite{Wei:2020ght}.
Examination of the innermost stable circular orbit and shadow \cite{Guo:2020zmf}.
Investigations into the thermodynamics, phase transitions, and Joule-Thomson expansion of (un)charged Anti-de Sitter black holes \cite{Hegde:2020xlv, Wei:2020poh, Wang:2020pmb}.
Exploration of Bardeen solutions \cite{Singh:2020xju}.
Research on rotating black holes \cite{Kumar:2020owy, Ghosh:2020vpc}.
Development of solutions for relativistic stars \cite{Doneva:2020ped}.
Investigation into Born-Infeld black holes \cite{Yang:2020jno}.
Examination of spinning test particles orbiting around static spherically symmetric black holes \cite{Zhang:2020qew}.
Studies on the thermodynamics and critical behavior of Anti-de Sitter black holes \cite{Ghosh:2020ijh}.
Exploration of gravitational lensing phenomena \cite{Islam:2020xmy}.
Analysis of the thermodynamic geometry of Anti-de Sitter black holes \cite{HosseiniMansoori:2020yfj}.
Investigations into Hayward black holes \cite{Kumar:2020xvu}.
Research on thin accretion disks around black holes \cite{Liu:2020vkh}.
Examination of superradiance and stability in charged Einstein-Gauss-Bonnet black holes \cite{Zhang:2020sjh}.
These studies collectively contribute to our understanding of various aspects of gravitational physics.

This paper delves into the realm of Einstein-Scalar-Gauss-Bonnet (ESGB) gravity involving a scalar field. The emergence of the  scalar field can occur through various means, including gravitational processes that arise from induced or spontaneous scalarization, or it can manifest as a gauge field through charged   solutions.  The presence of a scalar field has a notable impact on the gravitational backdrop and the structure of geodesics, which encompasses the trajectories of photons and the dimensions of a BH's shadow. Specifically, we will employ the field equations of ESGB theory that incorporate charge, alongside the cosmological constant. It's widely recognized that when the Gauss-Bonnet (GB) term is coupled in a non-minimal manner with a scalar field denoted as $\zeta$, the ensuing dynamics exhibit intricate and non-trivial characteristics See Â \cite{Nojiri:2005jg,Nojiri:2005am,Cognola:2006eg,Nojiri:2010oco,Cognola:2009jx,Capozziello:2008gu,Sadeghi:2009pu,Guo:2010jr,Satoh:2010ep, Nozari:2013wua,Lahiri:2016qih,Mathew:2016anx,Nozari:2015jca,Motaharfar:2016dqt,Carter:2005fu,Paolella,vandeBruck:2016xvt,Granda:2014zea,Granda:2011kx,Nojiri:2005vv,Hikmawan:2015rze,Kanti:1998jd,Easther:1996yd, Rizos:1993rt,Starobinsky:1980te,Mukhanov:1991zn,Brandenberger:1993ef,Barrow:1993hp,Damour:1994zq,Angelantonj:1994dv,Kaloper:1995tu,Gasperini:1996in,
Rey:1996ad,Rey:1996ka,Easther:1995ba,Santillan:2017nik,Bose:1997qv,KalyanaRama:1996ar,KalyanaRama:1996im,KalyanaRama:1997xt,Brustein:1997ny,Brustein:1997cv}
 and references therein.Nevertheless, there has been no prior exploration of charged spherically symmetric BH solutions within the ESGB   when adopting the assumption that the metric potentials do not follow the ansatz $ g_{tt}\neq g_{rr}{}^{-1}$. This paper endeavors to rectify this gap by obtaining precise spherically symmetric charged black hole solutions in the ESGB theory and subsequently delving into their physical attributes.
The  structure of the paper  is the following.
In Section \ref{S22}, we discuss the ghost-free EGB theory to investigate the formation of BH. Furthermore, we study  charged field equations to derive  exact  spherically symmetric space-times with  ${\color{blue} g_{tt}\neq g_{rr}{}^{-1}}$ .
In Sec.~\ref{S3}, we discuss  the  physics related to the  charged BH solutions, in particular,  the thermodynamical   properties.
In Sec.~\ref{S4}, we discuss  results and draw conclusions\footnote{The full length solutions are reported in a supplementary notebook.}

\section{Charged solutions in  Einstein-Scalar-Gauss-Bonnet gravity}\label{S22}
Within this section, we will provide a succinct overview of the development of ghost-free $f\left(\mathcal{G} \right)$ gravity employing Lagrange multipliers. Subsequently, we will proceed to get a solution characterized by spherically symmetric charge.

The Lagrangian for a ghost-free formulation can be expressed as follows, as detailed in \cite{Nashed:2021pah}:
\begin{align}
\label{FRGBg222}
\mathcal{L}=\int d^4x\sqrt{-g} \left[\frac{1}{2\kappa^2}R
+ \lambda \left( \frac{1}{2}\omega(\zeta) \partial_\mu \zeta \partial^\mu \zeta + \frac{\mu^4}{2} \right)
+ h\left( \zeta \right) \mathcal{G} - \tilde V\left( \zeta \right)
+ \mathcal{L}_\mathrm{matter}-\Lambda +\mathcal{L}_\mathrm{em}\right]\,.
\end{align}
In this equation, $\mathcal{G}$ represents the Gauss-Bonnet (GB) topological invariant, $V(\zeta)$ stands for the field potential, and $h(\zeta)$ is a function associated with the auxiliary field. Additionally, the constant $\mu$ carries a mass dimension.

The Lagrangian for the electromagnetic field, denoted as $\mathcal{L}_\mathrm{em}$, is defined as follows:
\begin{align}\mathcal{L}_\mathrm{em}= - \frac{1}{2}\mathcal{F}\;
\left(\mathcal{F}\equiv \mathcal{F}_{\mu \nu}\mathcal{F}^{\mu \nu}\right)\,.\end{align}
Here $\mathcal{F} = d\xi$ and
$\xi=\xi_{\beta}dx^\beta$ is the electromagnetic  
potential \cite{Awad:2017tyz,Nashed:2021pah}. See, in particular, Ref.\cite{Vasquez}.
Variations of the action (\ref{FRGBg222}) concerning $\zeta$,   $\lambda$, and $g_{\mu \nu}$ result in the following equations:
\begin{align}
 \label{FRGBg23gg}
0 =& - \frac{1}{\sqrt{-g}} \partial_\mu \left( \lambda \omega(\zeta) g^{\mu\nu}\sqrt{-g}
\partial_\nu \zeta \right)
+ h'\left( \zeta \right) \mathcal{G} - {\tilde V}'\left( \zeta \right) +\frac{1}{2} \lambda \omega'(\zeta) g^{\mu\nu}
\partial_\mu \zeta\partial_\nu \zeta\, ,\\
\label{FRGBg23}
 0=&\frac{1}{2} \omega(\zeta) \partial_\mu \zeta \partial^\mu \zeta + \frac{\mu^4}{2}\,,\\
\label{FRGBg24BB}
0 =& \frac{1}{2\kappa^2}\left(- R_{\mu\nu}
+ \frac{1}{2}g_{\mu\nu} [R-4\Lambda]\right) + \frac{1}{2} T^{\mathrm{matter}}_{\mu\nu}
 - \frac{1}{2} \lambda \omega(\zeta) \partial_\mu \zeta \partial_\nu \zeta
 - \frac{1}{2}g_{\mu\nu} \tilde V \left( \zeta \right)
+ D_{\mu\nu}^{\ \ \tau\eta} \nabla_\tau \nabla_\eta h \left( \zeta
\right)+T^{\mathrm{em}}_ {\mu\nu}\, .
\end{align}
The energy-momentum tensor of the electromagnetic field  $T^{\mathit{em}} _{\mu\nu}$, is built from  the Maxwell field as:
\begin{equation}
\label{en11}
T^{\mathrm{em}\, \nu}_{\mu}=\mathcal{F}_{\mu \alpha}\mathcal{F}^{\nu\alpha}
 -\frac{1}{4} \delta_\mu{}^\nu \mathcal{F}\, .
\end{equation}
Additionally, the variation of  Lagrangian~(\ref{FRGBg222}) w.r.t. the gauge potential 1-form
$\xi_{\mu}$ gives:
\begin{equation}
\label{q8b}
L^\nu\equiv \partial_\mu\left( \sqrt{-g} \mathcal{F}\mathcal{F}^{\mu \nu} \right)=0\, .
\end{equation}
In this paper, we  will not take into account  any more the  energy-momentum tensor of perfect fluid matter $\mathcal{L}_\mathrm{matter}$ because we  are interested only in vacuum solutions.

Now, let's examine the field equations (\ref{FRGBg23gg}), (\ref{FRGBg23}), (\ref{FRGBg24BB}), and \eqref{q8b} within a spherically symmetric spacetime characterized by non-equal metric potentials, where $g_{tt}\neq g_{rr}{}^{-1}$. Our goal is to solve these resulting   equations and obtain an analytical solution for the model.
The metric can be supposed of  the form:
\begin{equation}
\label{met}
ds^2 = f(r)dt^2 - \frac{dr^2}{f_1(r)} - r^2 \left[ d\theta^2 + \sin^2\theta \, d\phi^2\right]\,.
\end{equation}
Here, we have two unspecified functions of the radial coordinate denoted as $f(r)$ and $f_1(r)$. In our investigation, we make an assumption regarding the vector potential of the Maxwell field, as follows \cite{Nashed:2019zmy}:
\begin{align}
\label{vect} \xi=q(r)dt\,,
\end{align}
where $q(r)$ is the electric field which is a function of the radial coordinate.

From Eqs. (\ref{FRGBg23gg}), (\ref{FRGBg23}), (\ref{FRGBg24BB}),  and \eqref{q8b}   we obtain:
\begin{itemize}
\item the $(0,0)$  of Eq.~ (\ref{FRGBg23gg}) is:
\begin{align}
\label{eqtt}
& 0=-\frac {1}{f  {r}^{2}}\bigg[f    +12\, h' f'_1 f_1  f  -f\tilde V  {r}^{2} + q'^{2}f_1  {r}^{2}   -8h'' ff_1  +8h''ff_1{}^2 -4 h'ff'_1 - ff'_1 r  -ff_1 \bigg]\,,
\end{align}
\item the $(r,r)$ of Eq.~(\ref{FRGBg23gg}) is:
\begin{align}
\label{eqrr}
&0=\frac {1}{f  {r}^{2} \sin^2\theta}\bigg[\lambda  \omega  \zeta'^{2}f f_1  {r}^{2} \sin^2\theta+f\tilde V  {r}^{2} \sin^2\theta+4\, h'  f'f_1   \sin^2\theta-12\,h'f' f_1{}^{2} \sin^2\theta+f_1   f' r \sin^2\theta- q'^{2}f_1 {r}^{2} \sin^2\theta\nonumber\\
&+ff_1   \sin^2\theta -f   \sin^2\theta\bigg]\,,
\end{align}
\item the $(\theta,\theta)=(\phi,\phi)$  of Eq.~(\ref{FRGBg23gg}) is:
\begin{align}
\label{eqthth}
&0=-\frac {1}{4{r}^{2}f^{2} \sin^2\theta}\bigg[16\, h' f_1{}^{2}f  f'' r-2\,f_1  f''f {r}^{2}-8\, h' f_1{}^{2} f'^{2}r- f'_1 f' f {r}^{2}+16\, h'' f_1{}^{2} f'f r-2\,f_1 f' f r+24\, h'  f'_1 f_1  f' f r\nonumber\\
&-16\, h' f_1{}^{2}f  f'' r\cos^2\theta-16 \, h''f_1{}^{2} f' f r\cos^2\theta-24\, h' f'_1 f_1  f' f r\cos^2\theta+2\,f_1 f''f {r}^{2}\cos^2\theta+8\, h' f_1{}^{2} f'^{2}r\cos^2\theta\nonumber\\
&+2\,f_1  f' f r\cos^2\theta+4\,f f_1 {r}^{2} q'^{2}\cos^2\theta+ f'_1f'f {r}^{2} \cos^2 \theta -f_1  f'^{2}{r}^{2}\cos^2  \theta +4\,\tilde V  f'^{2}{r}^{2}\cos^2\theta+2 \, f'_1 f^{2}r\cos^2 \theta \nonumber\\
& -4\,f f_1 {r}^{2} q'^{2}-4\,\tilde V  f^{2}{r}^{2}-2\, f'_1 f^{2}r+f_1 f'^ {2}{r}^{2}\bigg]\,,
\end{align}
%
%
%
where $q'=\frac{\partial q(r)}{\partial r}$.
\end{itemize}
The field equations of the scalar field   (\ref{FRGBg23gg}) and \eqref{FRGBg23}  take the following forms:
\begin{align}
\label{eqphi11}
&0= \frac {1  }{2{r}^{2} f^{2}\zeta'}\bigg[12\, h' f'_1 ff' f_1-4\,h'f_1{}^{2} f'^{2}-2\, f^{2}{r}^{2} \zeta'^{2} \lambda'\omega  f_1  - f^{2}{r}^{2} \zeta'^{2}\lambda  \omega  f'_1 -4\, f^{2}r \zeta'^{2}\lambda  \omega  f_1 -2\,\tilde V' f^{2}{ r}^{2}-4\, h' f'_1 ff'\nonumber\\
& -8\, h'f_1  f''f  +8\, h'f_1{}^{2} ff'' +4\, h'f_1 f'^{2}-\lambda  \omega'  f_1   \zeta'^{2} f^{2 }{r}^{2}-f  {r}^{2} \zeta'^{2}\lambda  \omega f_1 f'-2\, f^{2}{r}^{2} \zeta' \lambda  \omega  f_1 \zeta''\bigg]
\,,
\end{align}
\begin{align}
\label{eqphi1}
0=\frac {2\,\lambda  \omega  f_1f  r \zeta''  +2\zeta'  \left\{\lambda'  \omega f_1 f  r+\left[\lambda   \omega'  ff_1  r+1/2\, \left( r ff'_1 +f_1  \left( 4\,b  +rf'   \right)  \right) \omega   \right] \right\} }{2b  r}\,.
\end{align}
Ultimately, the component of the field equations \eqref{q8b} that does not equate to zero yields the following result:
\begin{align}
\label{eqphie}
&L^t\equiv{\frac {rf_1 q'f'-r f'_1 q' f  -2\,rf_1 q'' f  -4\,f_1   q'f}{2 f^{2}r}}=0\,,
\end{align}
where $'=\frac{d}{dr}$, and  $''=\frac{d^2}{dr^2}$.

 Eqs.~(\ref{eqtt})-(\ref{eqphie}) are six non-linear differential equations for eight unknown functions $f$, $f_1$, $h$, $\tilde V'$, $\lambda$, $\omega$,  $\chi$, and $q$, therefore, we are going to fix some of these unknown functions to derive the other ones.
First, we solve Eq.~(\ref{FRGBg23}) and obtain\footnote{The special form of the solution of the scalar field given by Eq.~(\ref{scal}) does not allow us to put the constant $c_1$ equal zero, because in this case the function $\omega$ will be undefined. Moreover, if we assume the scalar field equal zero this yields the function $\omega$ will also vanishes and in that case the expression multiplied by the Lagrangian multiplier will be vanishing and the theory in that case will lost one of its main merits which is the ghost free.}
\begin{align}
\label{scal}
\zeta=c_1\,r\,, \qquad \qquad \omega(r)=-\frac{\mu^2}{c_1{}^2{ f_1(r)}}\,.
\end{align}
Using the above assumptions in Eqs. \eqref{eqtt}, \eqref{eqrr}, \eqref{eqthth}, \eqref{eqphi11} and \eqref{eqphi1},  we get:
\begin{align}
\label{sol4}
f(r)=\Lambda r^2+1+\frac{c_2r}{c_1+r^2}\,, \qquad \qquad f_1(r)=\Lambda r^2+\frac{c_2}{r}+\frac{c_1{}^{5/2}}{r^5}\,, \end{align}
while the forms of $h$, $\tilde{V}$ and $\lambda$ are reported in the Supplementary Material.  Let us analyze the above solution in which one can show that, from  the metric ansatzs, $f$ and $f_1$, cannot be equal to each other in any way. Moreover, the dimensional constant $c_1$, which has a dimension ${\textit L}^2$, is not allowed to be vanishing as Eq. (\ref{scal}) shows.

Due to the complicated forms of the electric field $q(r)$, the Lagrangian multiplier $\lambda$, the potential $V$, and  the arbitrary function $h$,  we are going to write here their  asymptotic  forms in view to understand analytically their behavior. The electric field $q(r)$,  for $r\to \infty$,  is:
\begin{align}\label{qs}
q(r)\approx c_3-\frac{c_4}{r}+\frac{c_4c_1c_2}{12\Lambda r^6}+\frac{c_2c_4(c_1+c_1c_2\Lambda -\Lambda)}{16\Lambda^2 r^8}-\frac{c_4c_1c_2{}^2}{18 \Lambda^2 r^9}\,.
\end{align}
Eq.~\eqref{qs} shows that if the constant $c_4$ is vanishing, we get a constant value of the electric field. Moreover Eq. \eqref{qs} shows that the electric field has more order than the monopole. These extra terms comes from the contributions due to the scalar field.
 The behavior of the potential $\tilde{V}(r)$ as $r$ approaches infinity can be expressed as follows:
\begin{align}\label{pot}
&\tilde{V}(r)\approx -3\Lambda -\frac{4\Lambda c_4{}^2}{3rc_2}+\frac{10c_1\Lambda}{3r^2}-\frac{20\Lambda c_1c_4{}^2}{9r^3c_2}-\frac{ 34\,{c_1}^{2}{\Lambda}}{9{{r}^{4}}}\,.
\end{align}
 Moreover, if the constant  $c_4$,  linked to the  electric, is vanishing, we see that the value of the potential  is  $O\left(\frac{1}{r^{2n}}\right)$. This implies that when there is no electric field present, the potential order follows a behavior of $O(\frac{1}{r^{2n}})$, where $n$ is a positive numerical value. Eq.~(\ref{pot}) indicates that we have a constant potential as $r\to \infty$.  Eq. (\ref{pot}) shows that the dimensional parameter $c_2$, which has the unit of ${\textit L}$, is not allowed to vanish to avoid undefined value of the potential $\tilde{V}$. Moreover, from Eq.~(\ref{qs}), we can not reproduce, in any case, the   Reissner Nordstr\"om gauge potential.

\begin{figure}
\centering
\subfigure[~${\tilde V}$ ]{\label{fig:v}\includegraphics[scale=0.3]{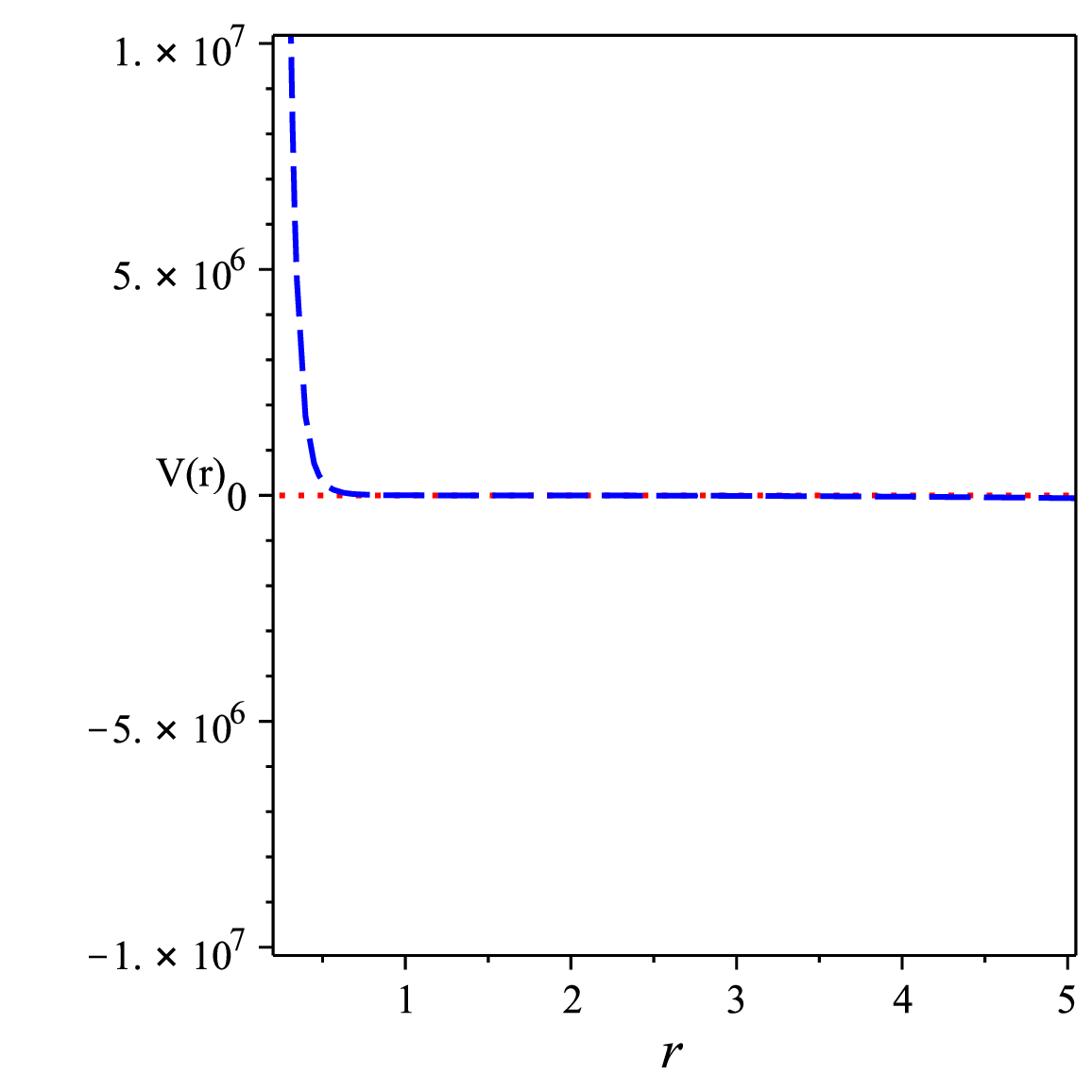}}
\subfigure[~ $h$]{\label{fig:h}\includegraphics[scale=0.3]{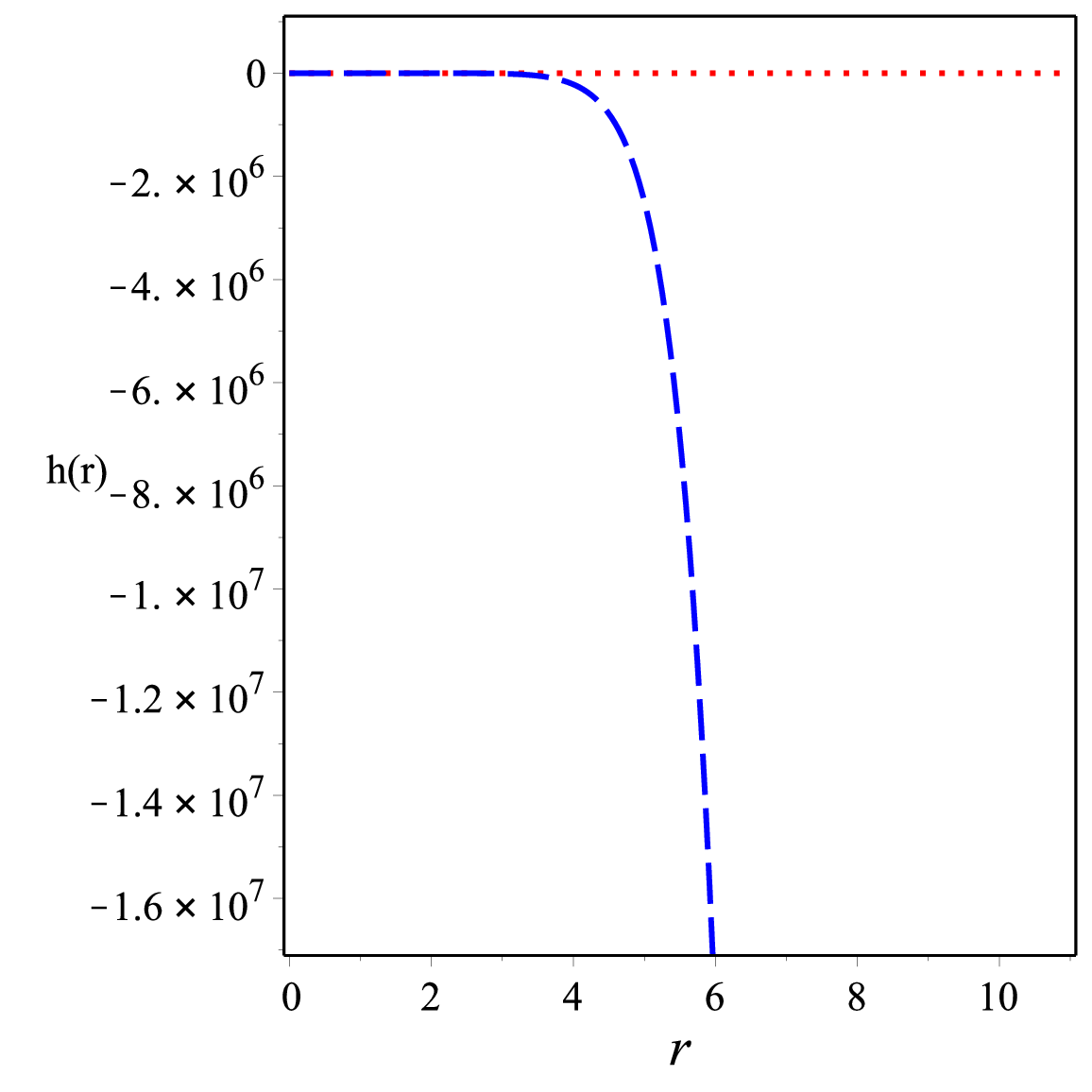}}
\subfigure[~$\lambda$]{\label{fig:l}\includegraphics[scale=0.3]{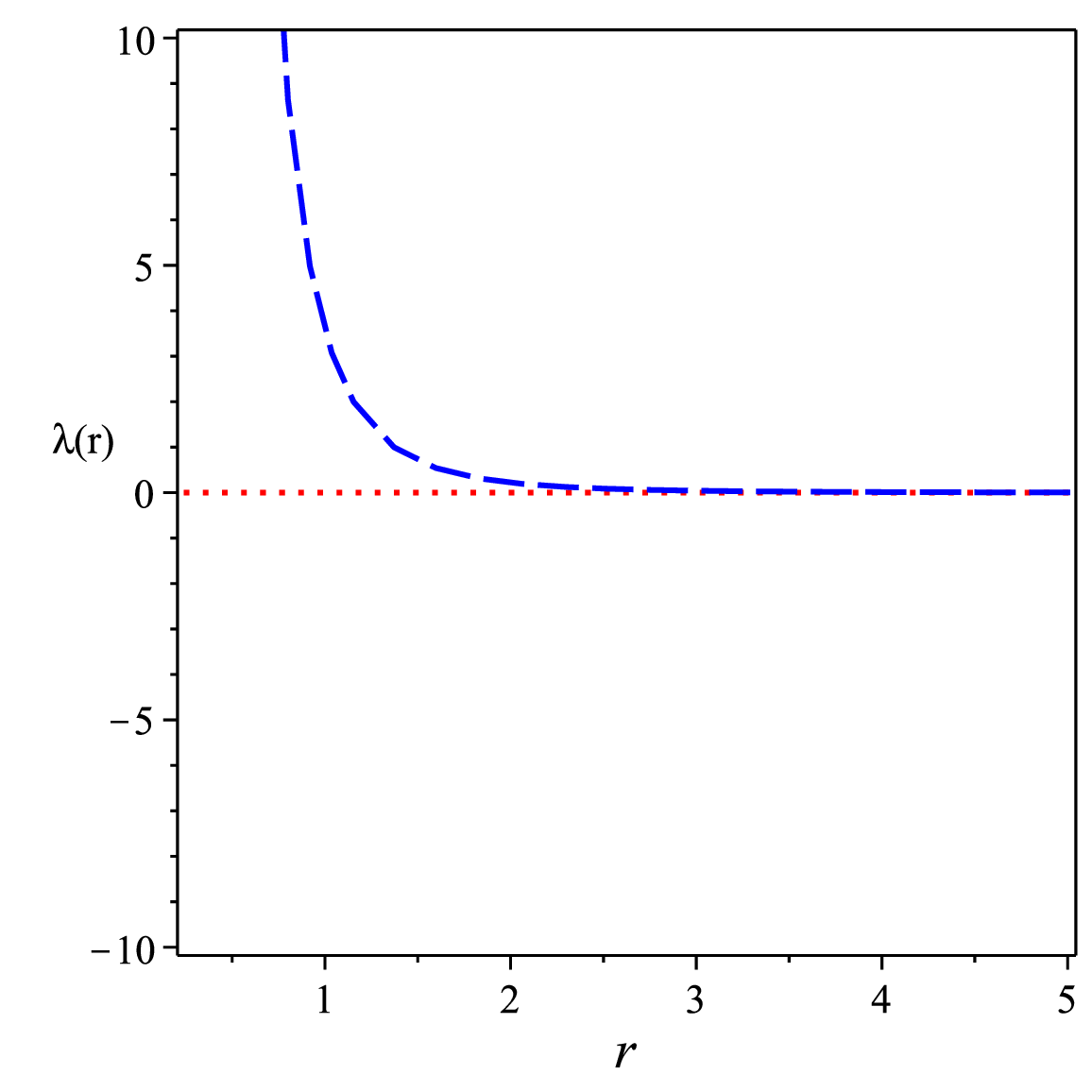}}
\caption[figtopcap]{\small{{
Plot \subref{fig:v} represents  ${\tilde V}$; Plot~\subref{fig:h} shows  the behavior of   $h$; Plot~\subref{fig:l} shows the behavior of $\lambda$. The numerical values of  parameters are $\Lambda=0.0001$,  $c_1=-5$,  and  $c_2=c_3=c_4=c_5=1$. The  dimensional constants are in  length units.}}}
\label{Fig:1}
\end{figure}

Now, let's examine the  trend exhibited by the function $h$ as it approaches zero. In this limit, it assumes the following form:
\begin{align}\label{has}
h(r)\approx c_5+\frac{r^7}{56 c_1{}^{7/2}}-\frac{c_3{r}^{10}}{60{c_1}^{5}} -\frac{r^{11}c_2(c_3{}^2+105)}{9240c_1{}^{6}}\,.
\end{align}
Eq.~ (\ref{has}) indicates that  $h$  is as $r\to0$.
Ultimately,  $\lambda(r)$, as $r\to \infty$, yields:
\begin{align}\label{alamb}
\lambda(r)\approx -\frac{4}3{\frac {{c_4}^{2}\Lambda}{c_2\,{\mu}^{4}r}}+\frac{10}3{
\frac {c_1\,\Lambda}{{\mu}^{4}{r}^{2}}}-{\frac {20}{9}}\,{\frac
{{c_4}^{2}\Lambda\,c_1}{c_2\,{\mu}^{4}{r}^{3}}}-{
\frac {1}{486}}\,{\frac {648\,{\Lambda}^{3}{c_4}
^{2}+486\,{\Lambda}^{3}c_1+1836\,{c_1}^{2}{
\Lambda}^{4}}{{\mu}^{4}{\Lambda}^{3}{r}^
{4}}}
\,.
\end{align}
Eq.~\eqref{alamb} means that, for $r \to \infty$,  we get a vanishing value of the Lagrangian multiplier. Also Eq.~\eqref{alamb} shows that if the constant $c_4$, related to the electric field, is vanishing, we get the behavior of the Lagrangian multiplier as  $O(\frac{1}{r^{2n}})$ with $n$ being positive number.
In Figure \ref{Fig:1} we indicate behavior of  ${\tilde V}$,  $h$ and   $\lambda$ for some numerical values  characterizing the solution. From these figures,  we can see that all these quantities have positive values and approach to zero as $r\to \infty$.

\section{Physical behavior of the charged black holes}\label{S3}
Now, we will delve into the physical characteristics of the black hole derived in the preceding section. To facilitate this, we express the line element of this black hole in the following manner:
\begin{align}\label{met11}
&ds^2=\bigg[\Lambda r^2+1+\frac{c_2r}{c_1+r^2}\bigg]dt^2-\frac{dr^2}{\Lambda r^2+1+\frac{c_2}{r}+\frac{c_1{}^{5/2}}{r^5}}-r^2(d\theta^2+\sin^2\theta d\phi^2)\nonumber\\
&\approx \bigg[\Lambda r^2+1-\frac{2M}r+\frac{2Mc_1}{r^3}-\frac{2Mc_1{}^2}{r^5}\bigg]dt^2-\frac{dr^2}{\Lambda r^2+1-\frac{2M}{r}+\frac{c_1{}^{5/2}}{r^5}}-r^2(d\theta^2+\sin^2\theta d\phi^2)\,.
\end{align}
 Here we put $c_2=-2M$. The above metric indicates that in no limit can we obtain a Reissner-Nordstr\"om BH.

 The curvature scalars, associated to the solution   (\ref{met11}), as $r\to \infty$ take the form:
 \begin{align}
 \label{inv}
 &R^{\alpha\beta \gamma \rho}R_{\alpha\beta \gamma \rho}\approx 24\Lambda^2-\frac{40Mc_1\Lambda}{r^5}+\frac{48M^2}{r^6}
 \,,\nonumber\\
 &R^{\alpha\beta}R_{\alpha\beta}\approx 36\Lambda^2+\frac{60Mc_1}{r^5}+\frac{24Mc_1-\Lambda c_1[252Mc_1-108] }{2r^7}
 \,,\nonumber\\
 &R\approx -12\Lambda-\frac{10Mc_1}{r^5}-\frac{2Mc_1+3\Lambda c_1[3c_1+14Mc_1]}{\Lambda r^7}
 \nonumber\\
 &\mathcal{G}\approx 24\Lambda^2+\frac{40M\Lambda c_1}{r^5}+\frac{12c_1}{r^6}
 \,.
\end{align}
For $r\to 0$, they are:
\begin{align}
 \label{inv1}
 &R^{\alpha\beta \gamma \rho}R_{\alpha\beta \gamma \rho}\approx C \,r^2+C_1\,r+C_2+\frac{C_3}r
 \,,\nonumber\\
 &R^{\alpha\beta}R_{\alpha\beta}\approx \frac{C_4}{r^{12}}+\frac{C_5}{r^{13}}+\frac{C_6}{r^{14}}
 \,,\nonumber\\
 &R\approx C_7\,r^2+C_8 r+C_9+\frac{C_{10}}{r}+\frac{C_{11}}{r^2}\,,
 \nonumber\\
 &\mathcal{G}\approx\frac{C_{12}}{c_1r^{11}}-\frac{C_{13}}{c_1r^{11}}-\frac{60Mc_1}{r^{13}}
 \,,
\end{align}
 where $C_i$, $i=0\cdots 13$ are constants constructed  from $c_1$ and $c_2$.
Eqs.~(\ref{inv}) and (\ref{inv1}) show that the above invariants become  infinite as $r\to 0$ and have constant values as $r \to \infty$. Moreover the invariants in Eq. \eqref{inv}  have a mild singularity compared with the Schwarzschild and  Reissner-Nordstr\"om  space-times \cite{Nashed:2018oaf}, i.e., the leading term for $r\to 0$ is ${\cal O}\left(\frac{1}{r^5}\right)$ while, in Schwarzschild  or in Reissner-Nordstr\"om space-time, it  is ${\cal O}\left(\frac{1}{r^6}\right)$. The main source of this mild singularity is the dimensional constant $c_1$ which is related to the Gauss-Bonnet scalar field.

To derive the horizon radii of solution (\ref{met11}), we have to solve  the equation $f_1(r)=0$  which gives seven roots. It is hard to derive the analytic expressions of such solutions but we  can deduce the asymptotic expressions and plot them using  some numerical values characterizing the model.  We discuss these solutions which depend on the sign of  $\Lambda$.\\

\subsection{ The case $\Lambda<0$}

\begin{figure}
\centering
\subfigure[~The metric  of Eq. (\ref{met11}) for $c_1=3$.]{\label{fig:met}\includegraphics[scale=0.3]{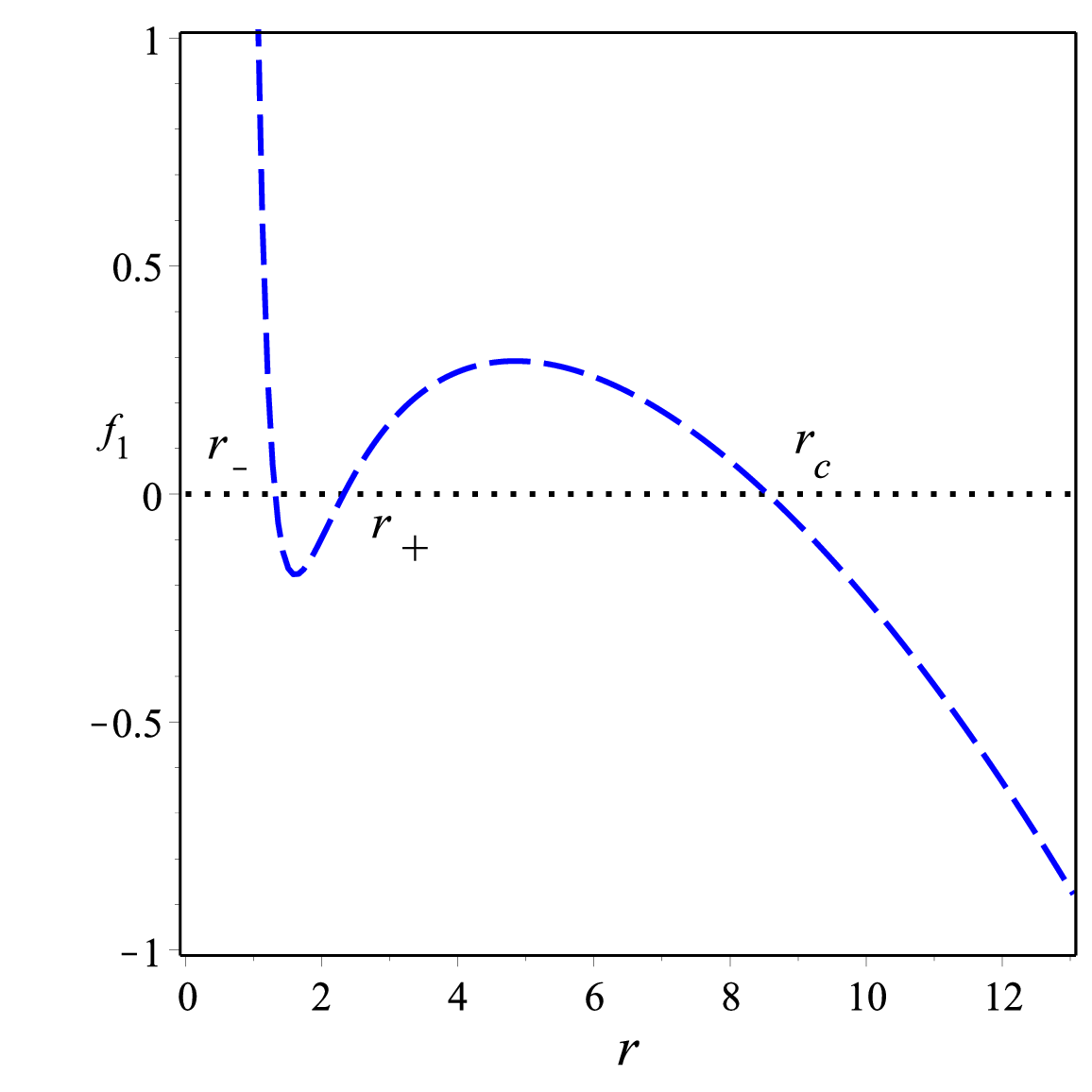}}
\subfigure[~The solutions  of Eq. (\ref{met11}) for $c_2=-2.3, -2.01$ and $c_2=-1.3$ respectively.]{\label{fig:met1p}\includegraphics[scale=0.3]{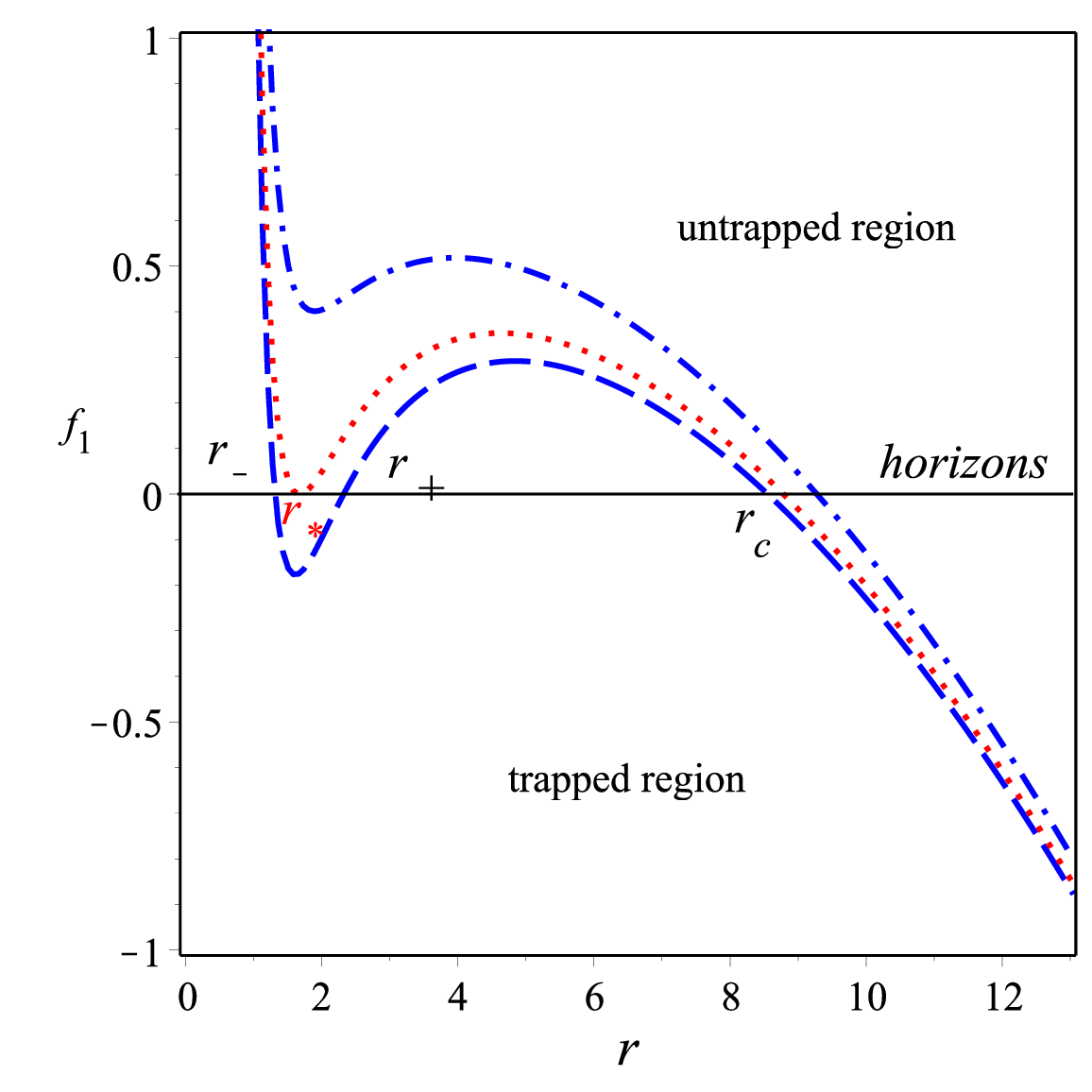}}
\subfigure[~The  temperature $T_+$ of Eq. (\ref{met11}) for $c_1=-1$.]{\label{fig:tempp}\includegraphics[scale=0.3]{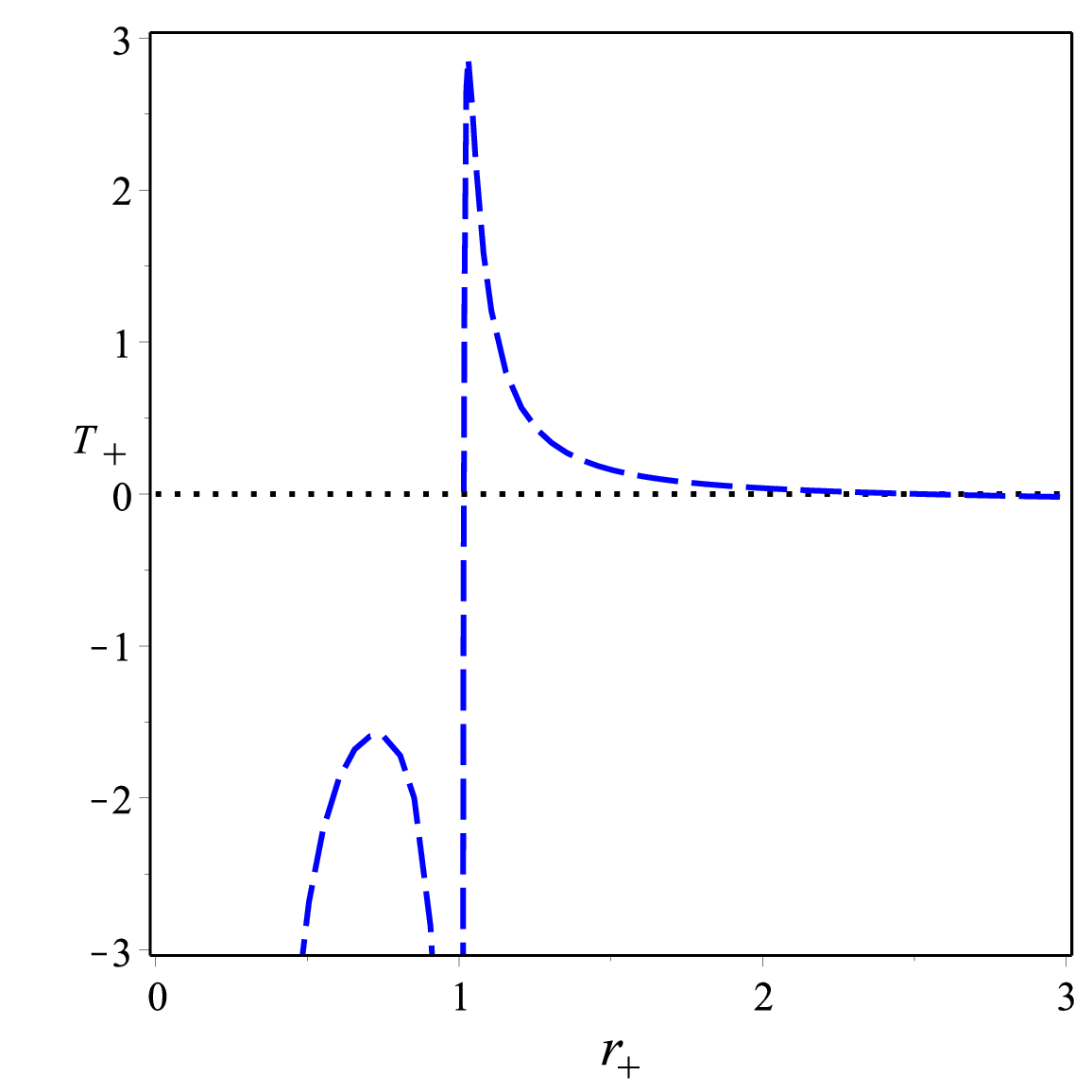}}
\subfigure[~The $H_+$ of Eq. (\ref{met11}) for $c_1=-1$.]{\label{fig:ent11p}\includegraphics[scale=0.3]{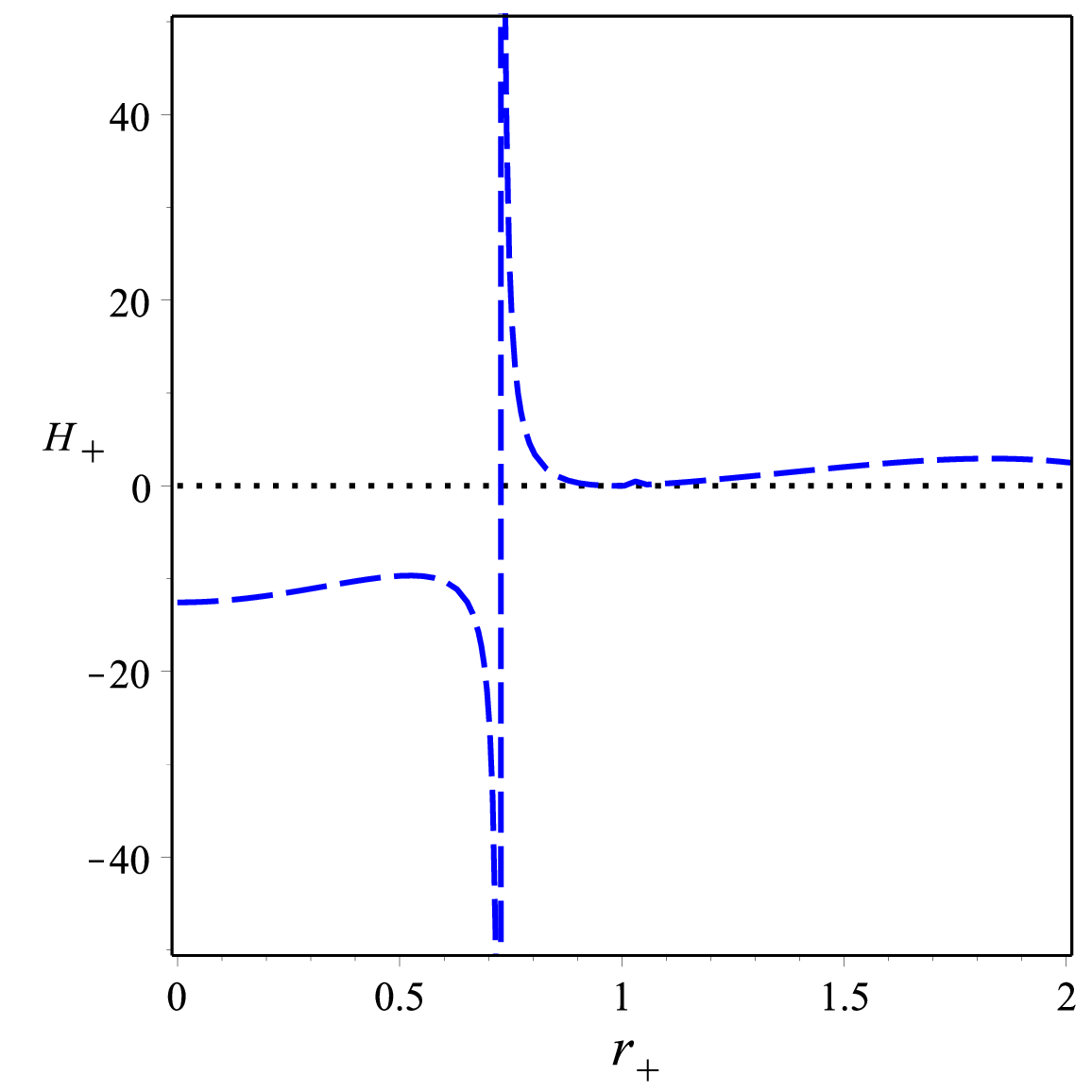}}
\subfigure[~The  $\mathbb{G}_+$ of Eq. (\ref{met11}) for $c_1=-1$.]{\label{fig:gib11p}\includegraphics[scale=0.3]{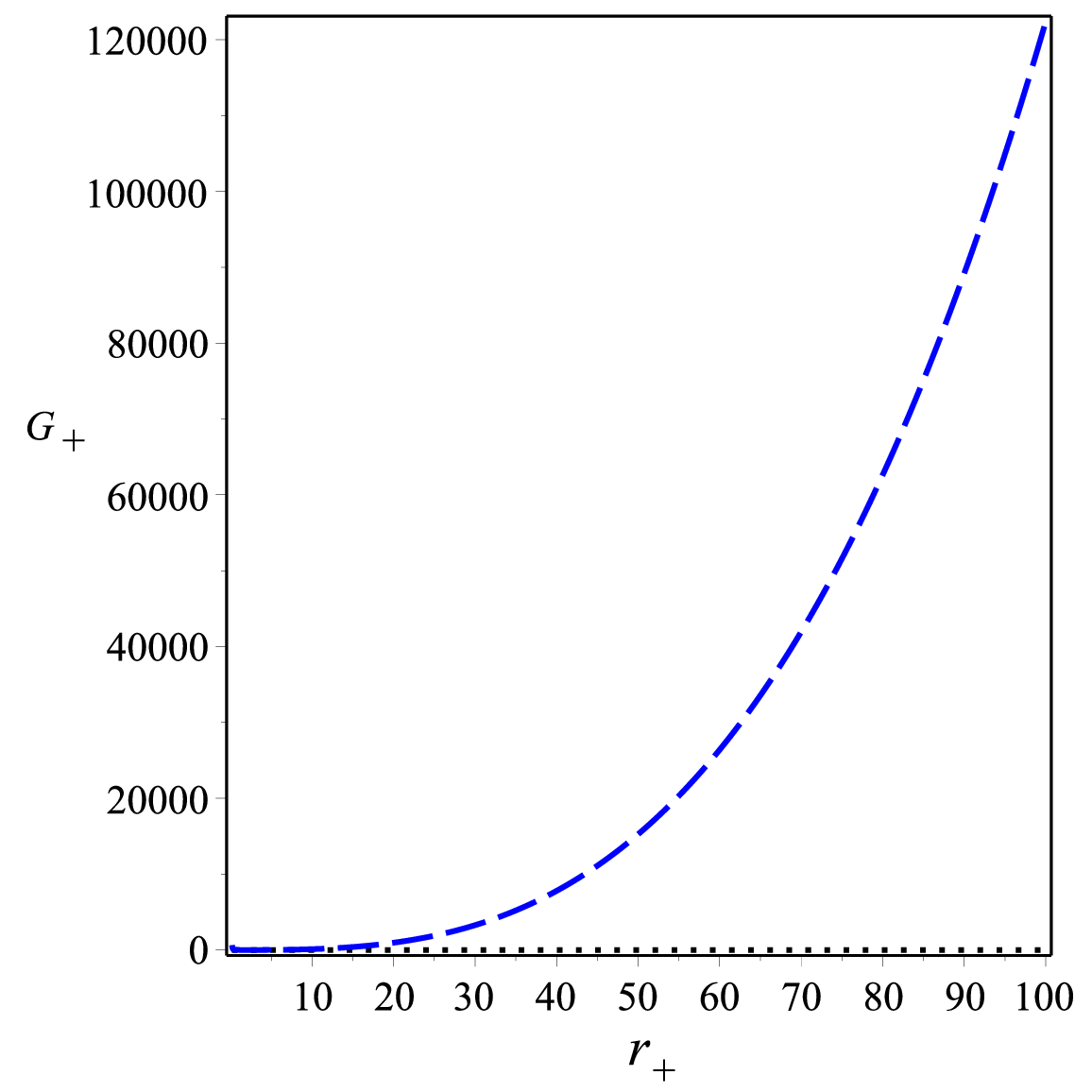}}
\caption[figtopcap]{\small{{Plot of the  thermodynamic quantities of  BH solution (\ref{met11}) for $\Lambda<0$;
Figure~\subref{fig:met} represents Eq. (\ref{met11}); Figure~\subref{fig:met1p} is the solutions of Eq. (\ref{met11});
Figure~\subref{fig:tempp} represents $T_+$;
Figure~\subref{fig:ent11p} represents $\mathbb{G}_+$; Figure~\subref{fig:gib11p} represents $\mathbb{G}_+$. Here we take  the numerical values for  $M=4.6$ and $\Lambda=-0.01$.}}}
\label{Fig:2}
\end{figure}

 Let us discuss now numerically the horizons of $g_{rr}{}^{-1}=f_1(r)=0$  since  analytic considerations are very difficult  being the  algebraic equation for $g_{rr}{}^{-1}=0$  of seventh order.  The solution   (\ref{met11})   has three real roots which represent the horizons for $g_{rr}{}^{-1}=f_1(r)=0$.   As shown in Figure~\ref{Fig:2}~\subref{fig:met},  this might constitute three, two, or one solutions depending on the relative values of $M$, $c_1$ and $\Lambda$. It is well-known that,  for charged  (A)dS solutions, one can derive two horizons. This model has  three horizons thanks to the dimensional parameter $c_1$ which  depends on the Gauss-Bonnet scalar field. Now we are going to discuss the formation of these three horizons separately:\\ i-When $\Lambda=-0.01$,  $c_1=3$,  and $M=0.65$ then the metric has one horizon  in the $(t,r,\theta, \phi$) coordinates. In this case,  the horizon emerges because of the cosmological constant $r_c$ and, if the  cosmological constant is equal zero, then we have a naked singularity where  the coordinate $t$ is always timelike and  $r$ is always spacelike. But still there is the singularity at $r = 0$, which is now  timelike. \\ ii--When $\Lambda=-0.01$,   $c_1=3$, and $M=1.01$,  the metric has two horizons  in the $(t,r,\theta, \phi$) coordinates. In this case,  $r_-$ and $r_+$  coincide to form $r_d$ which is the degenerate horizon and the other horizon is the one produced by the cosmological constant $r_c$. When the cosmological constant is vanishing, we get one horizon which occurs at $r=2M$.   This represents an event horizon, but the $r$ coordinate is never timelike: it becomes
null at $r=2M$, but it is spacelike on the other side. The singularity at $r =0$ is
timelike.\\iii-When $\Lambda=-0.01$,  $c_1=3$, and $M=1.15$,  the metric has three horizons  in the $(t,r,\theta, \phi$) coordinates. In this case, the metric is positive for large $r$ and small $r$ and negative inside the two vanishing points $r_\pm$ as Figure~\ref{Fig:2}~\subref{fig:met} shows. When the cosmological constant is equal zero,   the metric has a coordinate singularity at both $r_-$ and $r_+$ that could be removed by coordinate singularity. The surfaces defined by $r = r_\pm$ are both null, and they are both event horizons. The singularity at $r = 0$ is  timelike, not a spacelike surface as in
Schwarzschild. If one is an observer falling into the black hole from far away, $r_+$ is just like $2M$  in the Schwarzschild metric; at this radius $r$ switches from being a spacelike coordinate to a timelike coordinate, and one necessarily moves in
the direction of decreasing $r$.

We can discuss now  the thermodynamical properties of the BH solution (\ref{met11}).
To  this purpose,  let us  define some useful  expressions of thermodynamics. Figure~\ref{Fig:2}~\subref{fig:met1}, shows that  $r_-$ and $r_+$ are related to  the outer horizon as $r=r_+$ and the inner horizon of Cauchy as $r=r_-$, and
when  $c_1=3$, $\Lambda=-0.01$, and $c_2=2.3$ (trapped region). Furthermore, these horizons are related
and we can get a degenerate horizon $r_*$ (extreme BH)   when $c_2=-2.1$. If $c_2>-2$, there appears a naked singularity (untrapped region).

The   temperature $T$ of $r_+$ is figured as \cite{Sheykhi:2012zz,Sheykhi:2010zz,Hendi:2010gq,Sheykhi:2009pf,Wang:2018xhw,Zakria:2018gsf}:
\begin{align}
\label{temp}
T(r_+)= \frac{{\color{blue} f'}\left(r_+\right)}{4\pi}\,,
\end{align}
and the  entropy $S$   is figured as:
\begin{align}
\label{ent}
S\left(r_+\right) =\frac{1}{4} A \left(r_+ \right)\,,
\end{align}

The stability, $H$, of  the  BH  relies on the heat capacity sign.  To investigate this we figure $H_+$ as \cite{Nouicer:2007pu,Chamblin:1999tk}, 
\begin{equation}\label{m55}
H_+=\frac{\partial M}{\partial r_+} \left(\frac{\partial T}{\partial r_+}\right)^{-1}\,.
\end{equation}
When $H_{+} > 0\, (H_+ < 0)$,  the BH is    stable (unstable).
Additionally, we figure $\mathbb{G}\left(r_+\right)$ as \cite{Zheng:2018fyn, Kim:2012cma}:
\begin{align}
\label{enr}
\mathbb{G}\left(r_+\right)=M\left(r_+\right)-T\left(r_+\right)S\left(r_+\right)\,,
\end{align}
with $M\left(r_+\right)$  defined as:
\begin{align}\label{MM}
M\left(r_+\right)=\frac{\Lambda {r_+}^{7}+{r_+}^{5}+c_1}{2{r_+}^{4}}\,.
\end{align}
From Eq.~(\ref{temp}), we evaluate the Hawking temperature as:
\begin{align}
\label{T1}
T\left(r_+\right)=\frac{2\Lambda c_1{}^2  {r_2}^{5}+3c_1\Lambda {r_2}^7+3\Lambda r_+{}^9-c_1r_+{}^5+r_+{}^7-c_1{}^2+c_1r_+{}^2}{4\pi  r_+{}^4
(c_1+r_2{}^2)^{2}}\,.
\end{align}
The behavior of $H_{+}$  is represented in Figure~\ref{Fig:2}~\subref{fig:temp},  indicates that
$T\left(r_+\right) >0$ when  $r_+>r_*$.
Figure \ref{Fig:2}~\subref{fig:temp} indicates  that  $T_+$ vanishes as $r_2 = r_*$ and, if  $r_+<r_*$, we have a negative temperature. The vanishing of  temperature is a  critical temperature and thus we have a positive temperature above the critical temperature and  a negative temperature below the critical temperature.
We can calculate the heat capacity of  BH (\ref{met11}) obtaining:
\begin{align}
\label{en}
H\left( r_+ \right)=-\frac{2\pi(3\Lambda r_+{}^7+r_+{}^5-4c_1)(c_1+r_+{}^2)^3 }{2c_1{}^3\Lambda r_+{}^5+3c_1{}^2\Lambda r_+{}^7+12c_1\Lambda r_+{}^9+3\Lambda r_+{}^{11}-c_1{}^2r_+{}^5+6c_1r_+{}^7-r_+{}^9+6c_1{}^2r_+{}^2-6c_1{}^2r_+{}^4+4c_1{}^3}\, .
\end{align}
Using the above  thermodynamical quantities, we  can finally calculate  the  Gibbs function as:
\begin{align}
\label{enr}
\mathbb{G}\left(r_+\right)=\frac{-\Lambda c_1{}^2 {r_2}^{7}-11c_1\Lambda r_+{}^9-7\Lambda r_+{}^{11}-4c_1{}^2r_+{}^5-7c_1r_+{}^7-5r_+{}^9-4c_1{}^3-7c_1{}^2r_+{}^2-5c_1r_+{}^4}{4r_2{}^4(c_1+r_+{}^2)^2}\,.
\end{align}
The behavior of  $\mathbb{G}\left(r_+\right)$ is shown  in Figure~\ref{Fig:2}~\subref{fig:gib11} which  indicates its positivity.

\subsection{ The case $\Lambda>0$}

\begin{figure}
\centering
\subfigure[~The  BH of Eq. (\ref{met11}) when $M=4.6$, $c_1=3$, and $\Lambda=0.001$.]{\label{fig:met}\includegraphics[scale=0.3]{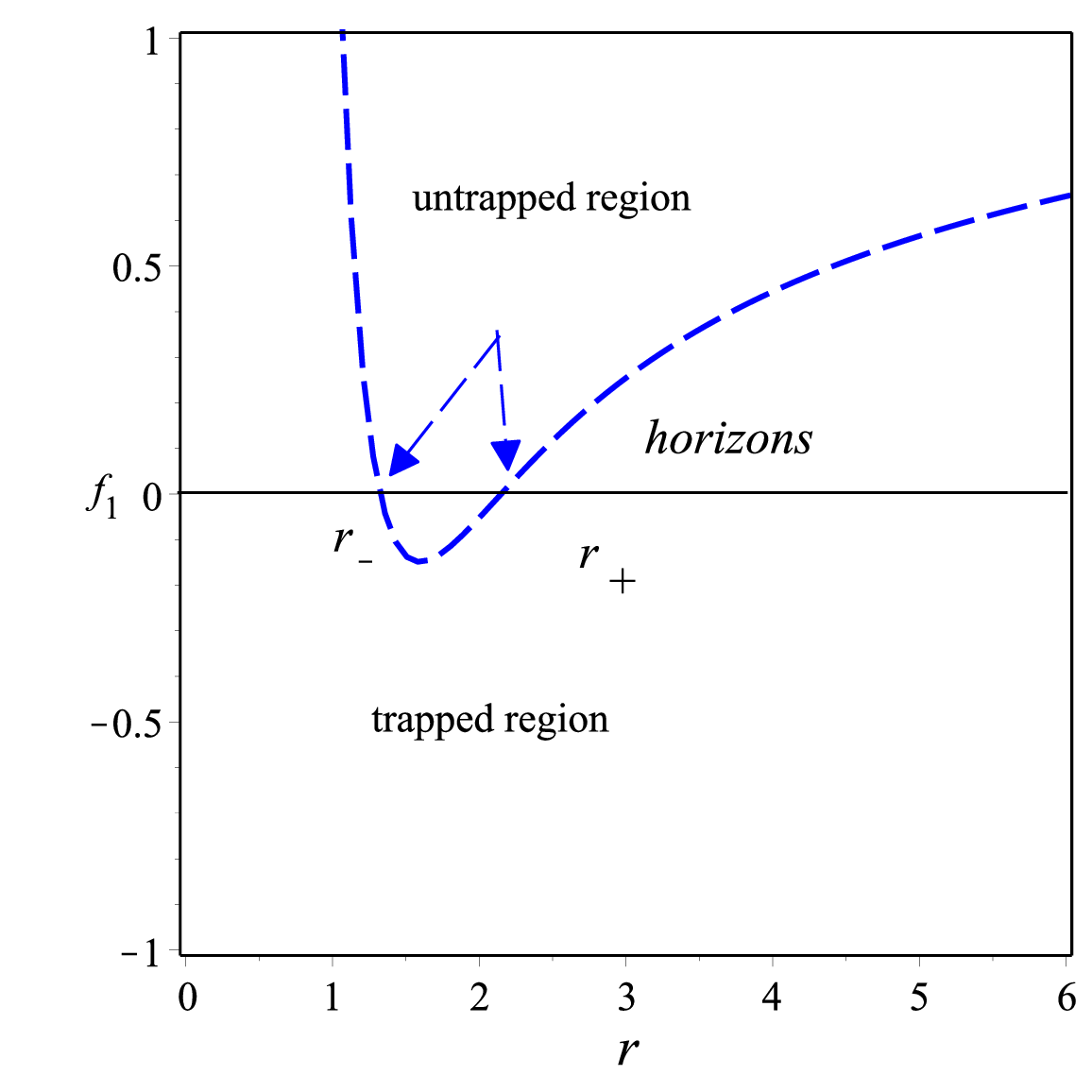}}
\subfigure[~The solutions of Eq. (\ref{met11}) when  $c_1=3$, $\Lambda=0.001$, $M=4.6$,  $M=4.12$, and  $M=2.6$, respectively.]{\label{fig:met1}\includegraphics[scale=0.3]{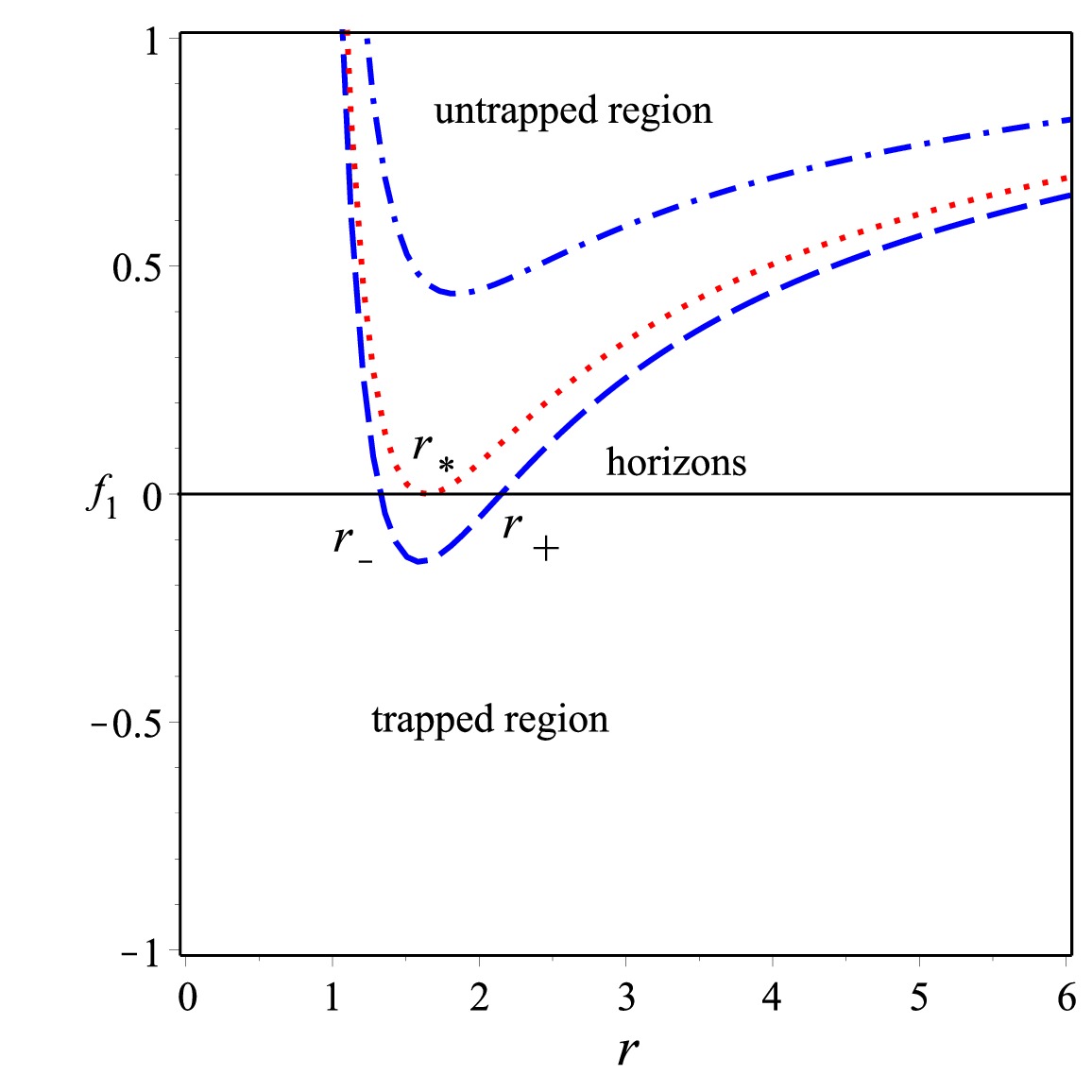}}
\subfigure[~The  temperature of Eq. (\ref{met11}), when $\Lambda=0.01$ and  $c_1=-0.3$.]{\label{fig:temp}\includegraphics[scale=0.3]{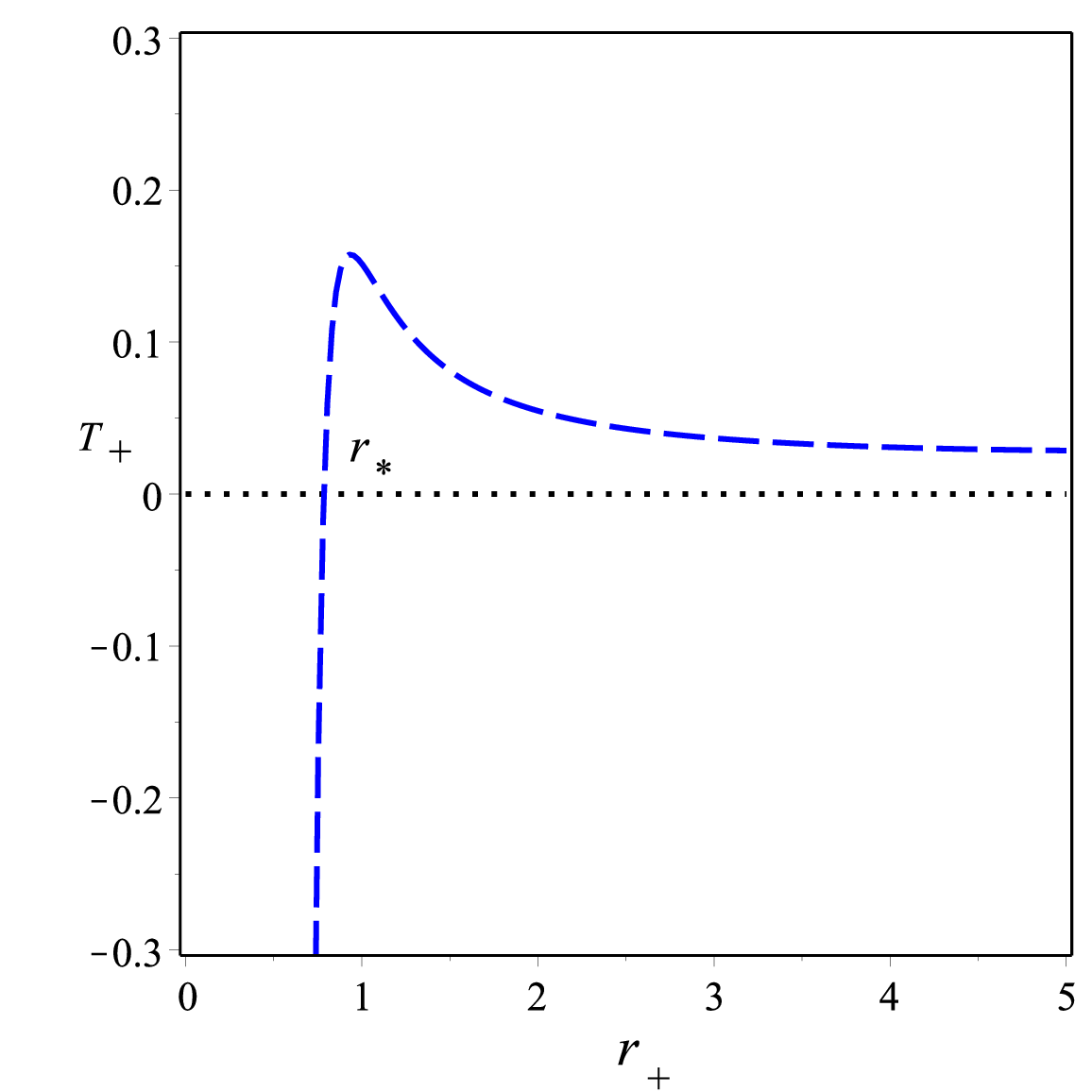}}
\subfigure[~The $H_+$ of BH (\ref{met11}) when $\Lambda=0.01$ and  $c_1=-0.3$.]{\label{fig:ent11}\includegraphics[scale=0.3]{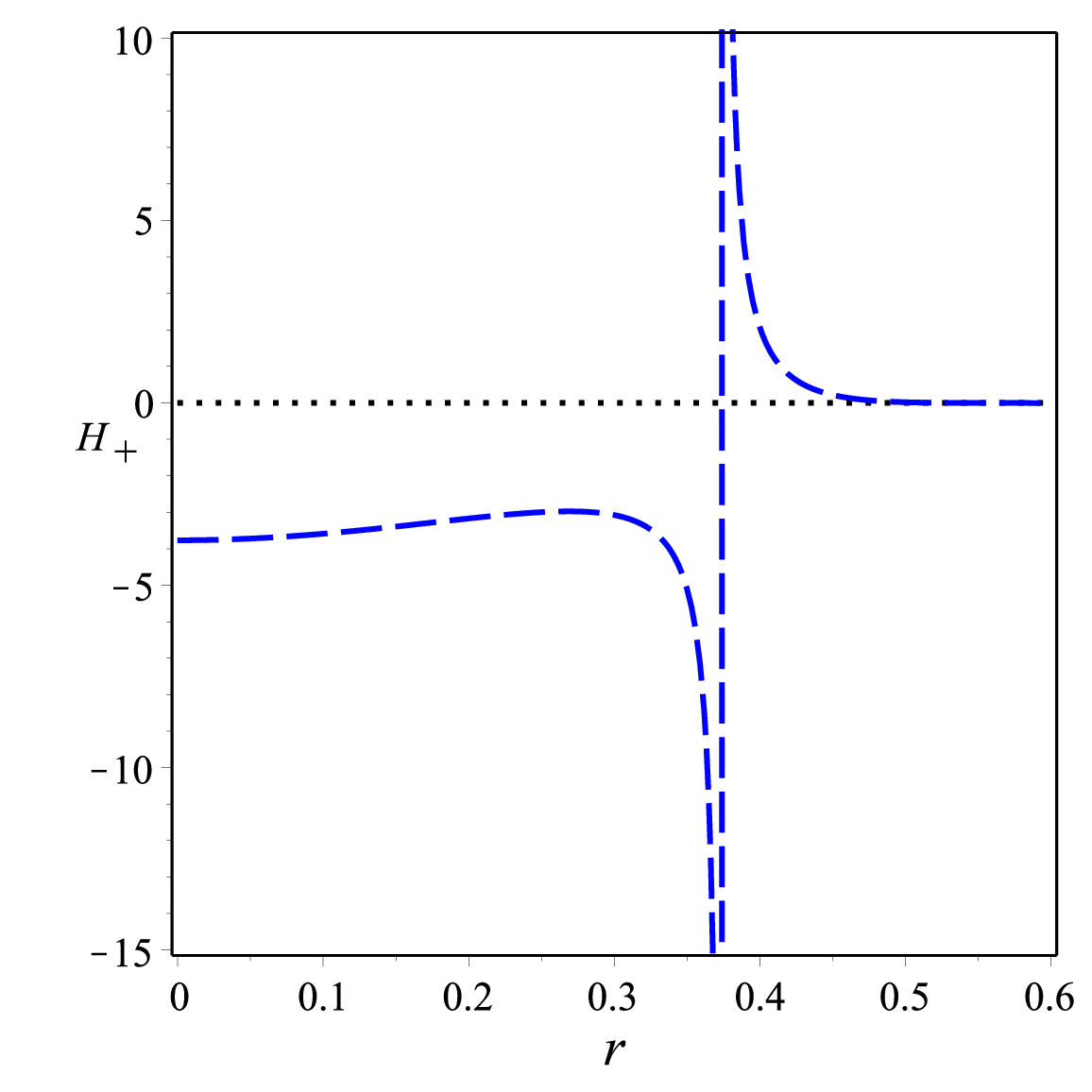}}
\subfigure[~The  $\mathbb{G}_+$ the BH (\ref{met11}) when $\Lambda=0.01$ and  $c_1=-0.3$.]{\label{fig:gib11}\includegraphics[scale=0.3]{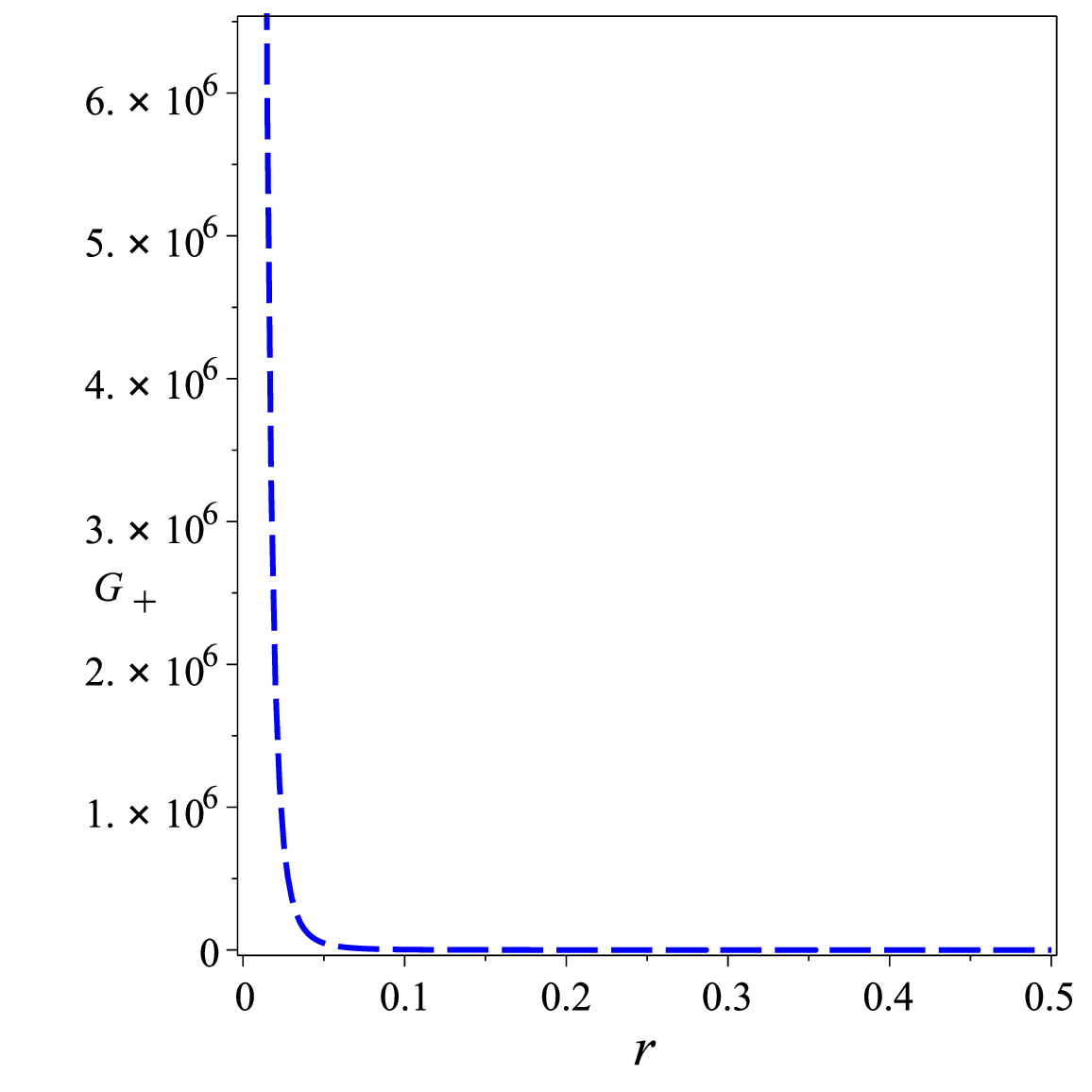}}
\caption[figtopcap]{\small{{Plot of the  thermodynamical quantities of Eq. (\ref{met11}) when $\Lambda>0$;
Figure~\subref{fig:met} represents Eq. (\ref{met11}); Figure~\subref{fig:met1} gives the solutions of Eq.(\ref{met11});
Figure~\subref{fig:temp} represents $T_+$;
Figure~\subref{fig:ent11} represents $H_+$; Figure~\subref{fig:gib11} represents $\mathbb{G}_+$.}}}
\label{Fig:3}
\end{figure}

Following the same procedure of the case $\Lambda<0$,  we obtain the behavior of the solutions of Eq. (\ref{met11}) for $\Lambda>0$. Also the behavior of the thermodynamical quantities like $T_+$, $H_+$  and $\mathbb{G}\left(r_+\right)$  are shown in Figure \ref{Fig:3}. From the figure,  it is clear that also for  $\Lambda>0$,  a physically relevant model can be achieved.

\section{Discussion and Conclusions}\label{S4}
 In four-dimension AdS space-time, we investigated a  class of GB gravitational equations coupled to a scalar field  \cite{Nojiri:2018ouv, Nojiri:2021mxf} in presence of the electromagnetic field. We obtained a   charged BH solution. It is derived  choosing the condition $ g_{tt}\neq g_{rr}{}^{-1}$  for the metric potentials. It is worth noticing that such a solution cannot yield the standard Reissner-Nordstr\"om space-time. The BH presents  extra terms  coming from the presence of the GB  field. Such a  contribution  is not allowed to have a zero value because one of the unknown functions  characterizing  the GB coupled scalar field, would be undefined if it vanishes. Therefore, this solution cannot, in general,  yield  solutions of Einstein's GR. Moreover, we have shown that the electric field of this solution is different from  the Maxwell field, and this difference depends on the contribution of the GB scalar field. This electric field cannot coincide with the electric field of Maxwell theory. Therefore, we can conclude that the GB scalar field, considered in this study, acts as a non-linear electrodynamics field reproducing higher multipole contributions  in the metric potential. In contrast to EGB theory in which one can not reproduce any solutions different from the solutions of GR since the GB term act sin this theory as  a topological invariant, which is constructed from the Riemann curvature tensor. Thus in the framework of Einstein theory, where the scalar field is coupled with the GB term, we succeeded in  deriving a new solution.

 We have to  stress again  the fact that the solution derived in this study can not be reduced, in any case,  to the Reissner-Nordstr\"om solution.  The reasons are the following:\\ $i)$ If the dimensional parameter $c_2$ is vanishing then we can reproduce  the Reissner-Nordstr\"om gauge potential but this give rise to  many issues in the solution as the potential field ${\tilde V}$  which is not defined (see Eq. (\ref{pot})). Additionally, being $c_2$ the total mass of the system, it can not go to zero otherwise the gravitational field has no source.

Furthermore, we investigated the relevant physics of the solution  and showed that, as $r \to \infty$, it  approaches to the AdS space-time. Furthermore, it is possible to  compute the associated curvature invariants  and  demonstrate that  the solution has a soft singularity when compared to the Schwarzschild and Reissner-Nordstr\"om space-times. This is because  the invariants of  Schwarzschild Â space-time behave as $\{K,R^{\mu \nu}R_{\mu \nu}, R\}\approx \{{\cal O}\left(\frac{1}{r^6}\right), 0,0\}$ while the invariants of this BH  behave as $\{K,R^{\mu \nu}R_{\mu \nu}, R\}\approx \{{\cal O}\left(\frac{1}{r^5}\right), {\cal O}\left(\frac{1}{r^5}\right),{\cal O}\left(\frac{1}{r^5}\right)\}$.  The main source of this soft singularity comes from  the Gauss-Bonnet scalar field.  Also, it is possible to show that this BH has  a multi-horizon structure. Specifically, we have  three-horizons for the AdS space-time and  two horizons for the  dS space-time.

Finally,  the thermodynamic behavior of this solution can be studied by Â calculating some  Hawking temperature, heat capacity, and Gibbs free energy. Thermodynamics can be divided into two classes depending on  cosmological constant sign, i.e., if $\Lambda>0$ or $\Lambda<0$. For both classes, we show that, below a critical temperature, Â the BH solution is unstable and undergoes a phase transition whose endpoint is the charged BH dressed with a scalar field. The solution  shows a second-order phase transition in which the scalar field condenses below a critical temperature. The Gibbs function can be also evaluated and we showed that, for both classes,  it always possesses a positive value.

To summarize, we found a charged solution in the framework of GB gravity coupled to a scalar and demonstrated that it evolves asymptotically  to the AdS space-time  as $r \to \infty$. Its singularity is soft compared to the Schwarzschild space-time,  it has a multi-horizons structure, and that it is unstable at the critical temperature.

A discussion similar to that reported here  can be developed for    GB gravity coupled to a scalar  but in  the framework of non-linear electrodynamics. We expect  different  BH solutions. This topic will be developed in a forthcoming  paper.

\acknowledgments
S.C. acknowledges the support of Istituto Nazionale di Fisica Nucleare, Sezione di Napoli, {\it iniziative specifiche} QGSKY and MOONLIGHT2. This paper is based upon work from COST Action CA21136 - \textit{Addressing observational tensions in cosmology with systematics and fundamental physics} (CosmoVerse), supported by COST (European Cooperation in Science and Technology).

\end{document}